\documentclass[12pt]{article}
\usepackage{graphicx}
\usepackage{amsfonts}
\usepackage{amssymb}
\usepackage{graphics,psboxit,amsmath}
\usepackage{subfigure}
\usepackage{verbatim}
\usepackage{epic}

\def\ba{\begin{equation}}
\def\ea{\end{equation}}

\newcommand{\be}{\begin{eqnarray}}
\newcommand{\ee}{\end{eqnarray}}
\newcommand{\non}{\nonumber}

\begin{document}

\begin{titlepage}
\begin{center}
\strut\hfill
\vskip 1.3cm


\vskip .5in

{\Large \bf The sine-Gordon model in the presence of defects}

\vskip 0.5in

{\bf Jean Avan$^{a}$  and Anastasia Doikou$^{b}$}
\vskip 0.2in

\noindent
{\footnotesize  $^a$LPTM, Universite de Cergy-Pontoise (CNRS UMR 8089),
F-95302 Cergy-Pontoise, France}\\
\vskip .1in
{\footnotesize $^{a}$Department of Engineering Sciences, University of Patras,
GR-26500 Patras, Greece}
\vskip .1in


{\footnotesize {\tt E-mail: avan@u-cergy.gr, adoikou@upatras.gr}}\\

\end{center}

\vskip 1.0in

\centerline{\bf Abstract}
The sine-Gordon model in the presence of dynamical integrable defects is investigated.
This is an application of the algebraic formulation introduced for integrable defects in earlier
works. The quantities in involution as well as the associated Lax pairs are explicitly extracted.
Integrability i also shown using certain sewing constraints, which emerge as suitable continuity
conditions.

\end{titlepage}
\vfill \eject

\section{Introduction}

The investigation of integrable defects has
been a quite challenging problem, and there is a wealth of relevant articles in recent years at
both classical and quantum level \cite{avan-doikou-defect}--\cite{agui}.
A fully algebraic picture for a description of a Liouville
integrable defect was recently proposed \cite{avan-doikou-defect}. It was exemplified in the case of the continuous non-linear
Schr\"{o}dinger model (NLS) \cite{avan-doikou-defect}, and now the procedure is extend to the sine Gordon model \cite{avan-doikou-defect2}.

This procedure will now be applied to the sine-Gordon (SG) model, for which we shall consider
a generic dynamical defect (type II) (see also \cite{fus, BCZ3}).
We must immediately emphasize that this model provides an example where the initial off-shell
continuous ``conserved'' Hamiltonians do not Poisson commute, but will be shown to weakly
Poisson-commute once the sewing conditions are implemented, thereby guaranteeing Liouville
integrability of the reduced model. By contrast in the NLS case the continuous Hamiltonians
strongly (i.e. off-shell) Poisson-commuted. As seen above this does not modify the conclusions on
Liouville integrability on-shell.

This paper is mainly based on the analysis of \cite{avan-doikou-defect2}, therefore for more details on the explicit algebraic formulation we refer
the interested reader to the aforementioned relevant article.

\section{General setting}

The key point in the description of classical integrable lattice models is the existence of the Lax pair ${\mathbb U},\ {\mathbb V}$. Define $\Psi$ as being a solution of the following set of equations (see e.g. \cite{ftbook})
\be
&&{\partial \Psi \over \partial x} = {\mathbb U}(x,t, \lambda)  \Psi \label{dif1}\\
&& {\partial \Psi \over \partial t } = {\mathbb
V}(x,t,\lambda) \Psi \label{dif2}
\ee
${\mathbb U},\ {\mathbb V}$ being in general $n \times n$ matrices with entries defined as
functions of complex valued dynamical fields, their derivatives, and the complex spectral parameter $\lambda$.
Compatibility of the two aforementioned equations (\ref{dif1}), (\ref{dif2}) gives rise to the zero curvature condition
\be
\dot{ {\mathbb U}}(x,t) -{\mathbb V}'(x,t) + \Big [ {\mathbb U}(x,t),\ {\mathbb V}(x,t)\Big ]=0,
\ee
which provides the equations of motion of the system at hand.

As is well known the generating function of the local integrals of motion is given by the expression
\be
{\cal G} = \ln (tr T(L, -L,\lambda)),
\ee
where the monodromy matrix $T$ is defined as,
\be
T(L, -L, \lambda)=P\exp \Big \{\int_{-L}^L dx\ {\mathbb U}(x) \Big \},
\ee

We now impose that the operator ${\mathbb U}$ satisfy the ultra-local Poisson structure described
by the linear algebraic relations
\be
\Big \{{\mathbb U}_1(x,\lambda),\ {\mathbb U}_2(y, \mu)\Big \} = \Big  [r_{12}(\lambda -\mu),\ {\mathbb U}(x, \lambda)+
{\mathbb U}_2(y, \mu)\Big ] \delta(x-y)\label{linear}
\ee
It is then straightforward to show that $T$ satisfies the fundamental quadratic algebra:
\be
\Big \{ T_1(\lambda),\ T_2(\mu) \Big \} = \Big [r_{12}(\lambda -\mu),\ T_1(\lambda)\ T_2(\mu) \Big ].  \label{fundam}
\ee
$r_{12}(\lambda -\mu)$ is the so-called classical $r$-matrix assumed here to be a non-dynamical skew-symmetric solution of the classical Yang-Baxter equation.

Let us now particularize our analysis to the sine-Gordon model. In this case
the ${\mathbb U}$ operator of the Lax pair is a $2 \times 2$ matrix and is given by \cite{ftbook}:
\be
{\mathbb U}(x, t, u) = {\beta \over 4i} \pi(x,t)\sigma^z +{mu \over 4i}e^{{i\beta \over 4}\phi\sigma^z} \sigma^y e^{-{i\beta \over 4}\phi\sigma^z}-
{mu^{-1} \over 4i}e^{-{i\beta \over 4}\phi\sigma^z} \sigma^y e^{{i\beta \over 4}\phi\sigma^z}
\ee
$u \equiv e^{\lambda}$, $\sigma^{x, y, z}$ are the $2\times 2$ Pauli matrices, and the associated classical
$r$-matrix in this case is given by the familiar form \cite{ftbook}:
\be
r(\lambda) = {\beta^2 \over 8 \sinh \lambda }\begin{pmatrix}
{\sigma^z +1\over 2} \cosh \lambda & \sigma^-\\
\sigma^+ & {-\sigma^z +1\over 2} \cosh \lambda \end{pmatrix}.\label{rm1}
\ee
Stating that the Lax operator ${\mathbb U}$ satisfies the linear Poisson algebra (\ref{linear})
is equivalent to setting that $\phi,\ \pi$ are canonical conjugates, i.e.
\be
\Big \{ \phi(x),\ \pi(y)\Big \}= \delta(x-y).
\ee

Let us now apply the generic defect construction to the sine-Gordon model.

\section{The sine-Gordon model with defect}

The main ingredient in the description of integrable defects is the modified monodromy matrix.
More precisely, in the presence of an integrable defect the monodromy matrix of the field theory
is modified (see also \cite{BCZ1, haku, avan-doikou-defect}), and takes the generic form
\be
T(L, -L, \lambda)&=& T^+(L, x_0^+, \lambda)\ \tilde L( \lambda)\ T^-(x_0^-, -L,\lambda)\non\\ &=& P\exp \Big \{\int_{x_0^+}^L dx\ {\mathbb U}^+(x) \Big \}\
\tilde L(\lambda)\ P\exp\Big \{\int_{-L}^{x_0^-}dx\ {\mathbb U}^-(x) \Big \} \label{contmon}
\ee

Assuming that the defect Lax matrix $\tilde L$ also satisfies the quadratic Poisson algebra
(\ref{fundam}) $T$ given in (\ref{contmon}) satisfies (\ref{fundam}).

A consistent generic parametrization of an integrable defect of the so-called Type-II or dynamical
will be considered. We assume the following form for the classical $\tilde L$ matrix
\be
\tilde L(\lambda) = \begin{pmatrix}
e^{\lambda}V -e^{-\lambda}V^{-1} & \bar a\\
a & e^{\lambda}V^{-1} -e^{-\lambda}V
\end{pmatrix}. \label{LI}
\ee

Requiring that $\tilde L$ satisfies the algebraic relation (\ref{fundam}), one then
extracts the following Poisson relations between the defect fields:
\be
&& \Big \{V,\ \bar a \Big \} = {\beta^2 \over 8 }V\ \bar a, \non\\
&& \Big \{V,\ a\Big \} = -{\beta^2 \over 8 }Va, \non\\
&& \Big \{ \bar a,\ a\Big \} = {\beta^2 \over 4 } (V^2 - V^{-2}) \label{dalg}
\ee

From these Poisson brackets one naturally extracts a cyclic variable $C_0 = V^2 + V^{-2} + \bar a a $
identified as the Casimir of a deformed $\mathfrak{sl}_2$. This variable Poisson-commute with all other dynamical
quantities and can therefore be fixed to some particular value $c_0$. We shall nevertheless
keep the redundant three-parameter expression for $\tilde L$ for reasons of form simplicity
in the explicit expressions.

We shall express the term of order $u$ in ${\mathbb U}$
independently of the fields, after applying a suitable gauge transformation \cite{ftbook}
\be
T^{\pm}(x,y,\lambda)= \Omega^{\pm}(x)\ \tilde T^{\pm}(x,y)\ (\Omega^{\pm}(y))^{-1}, ~~~~~\Omega^{\pm} = e^{{i\beta \over 4}\phi^{\pm} \sigma^z},
\ee
The gauge transformed operator $\tilde {\mathbb U}$ is expressed as:
\be
\tilde {\mathbb U}^{\pm}(x,t,u)=  {\beta \over 4i} \mathfrak{f}^{\pm} \sigma^z + {mu \over 4i} \sigma^y -{mu^{-1}\over
4i}e^{-{i \beta \over 2}\phi^{\pm} \sigma^z} \sigma^y e^{{i \beta \over 2 }\phi^{\pm} \sigma^z}
\ee
where we define
\be
\mathfrak{f}^{\pm}(x,t) = \pi^{\pm}(x,t) + \phi^{\pm'}(x,t).
\ee

We consider the following convenient decomposition for $\tilde T$, as $|u| \to
\infty$ \cite{ftbook},
\be
\tilde T^{\pm}(x,y,\lambda) = (1 +W^{\pm}(x, \lambda))\ e^{Z^{\pm}(x,y ,\lambda)}\
(1 +W^{\pm}(y, \lambda))^{-1} \label{expa}
\ee
$W^{\pm}$ is an off-diagonal
matrix and $Z^{\pm}$ is purely diagonal. They are expanded as:
\be
W^{\pm} = \sum_{k=0}^{\infty} {W^{\pm(k)} \over u^k},
~~~~~Z^{\pm} = \sum_{k=-1}^{\infty}{Z^{\pm(k)} \over u^k} \label{expa1}
\ee
Note that $\tilde T$ naturally satisfies the gauged Lax equation:
\be
{\partial \tilde T^{\pm} \over \partial x} =
\tilde {\mathbb U}^{\pm}(x, \lambda) \tilde T^{\pm}(x, y, \lambda) \label{dif1b}
\ee

Inserting expressions (\ref{expa}), (\ref{expa1}) in
(\ref{dif1b}) one identifies the matrices $W^{\pm(k)}$ and $Z^{\pm(k)}$. More precisely, we end up with
an equation for the off-diagonal matrix:
\be
{\partial W^{\pm} \over \partial x} + W^{\pm} \tilde {\mathbb U}^{\pm}_D - \tilde {\mathbb U}^{\pm}_D W^{\pm} + W^{\pm} \tilde {\mathbb U}^{\pm}_A W^{\pm} -{\mathbb U}^{\pm}_A =0
\ee
where the indices $D,\ A$ denote the diagonal and anti-diagonal part of the Lax operator $\tilde {\mathbb U}^{\pm}$,
leading to typical Riccati-type equations for the entries of $W^{\pm}$ \cite{avan-doikou-defect2}.

Similarities with corresponding equations emerging in \cite{caudr, agui} from the inverse scattering point
of view are apparent as expected, given that one essentially solves the same fundamental equations.
Liouville integrability is guaranteed within the
present approach by construction, at least formally,
whereas in the methodology of \cite{caudr, agui} only the conservation of the charges for a singled-out
time-evolution is shown through the zero curvature condition i.e. explicit use of the equations of motion.

It is sufficient for our purposes here to identify only the first few terms of
the expansions. Indeed based on equation (\ref{dif1b}) we conclude (see also \cite{ftbook}):
\be
&& W^{\pm(0)} = i \sigma_1, ~~~~W^{\pm(1)} = -{i \beta \over  m} \mathfrak{f}^{\pm}(x) \sigma_1, \non\\
&& W^{\pm(2)} = {2i \beta \mathfrak{f}^{\pm'}\over m^2}\ \sigma_2 - i \sin (\beta \phi^{\pm})\ \sigma_2 -
{\beta^2 (\mathfrak{f}^{\pm})^2\over 2im^2}\ \sigma_1.
\ee
We also need to identify the diagonal elements $Z^{\pm(n)}$.
In particular from equation (\ref{dif1}) we extract the following expressions:
\be
&& Z^{+(-1)}=  -{i m (L-x_0)\over 4} \sigma_3,  ~~~~~Z^{-(-1)}= -{i m (L+ x_0)\over 4} \sigma_3\non\\
&& Z^{+(1)} = { m \over 4} \begin{pmatrix}
  -\int_{x_0^+}^L dx\ W_{21}^{+(2)}(x) &        \\
     & \int_{x_0^+}^L dx\ W_{12}^{+(2)}(x)
\end{pmatrix} \non\\ &&- {m\over 4} \begin{pmatrix}
  -i \int_{x_0}^L dx\ e^{-i \beta \phi^+}  &        \\
     & i \int_{x_0^+}^L dx\ e^{i \beta \phi^+}
\end{pmatrix},  \non\\
&& Z^{-(1)} = { m \over 4} \begin{pmatrix}
  -\int_{-L}^{x_0^-} dx\ W_{21}^{-(2)}(x) &        \\
     & \int_{-L}^{x_0^-} dx\ W_{12}^{-(2)}(x)
\end{pmatrix} \non\\&& - {m\over 4} \begin{pmatrix}
  -i \int_{-L}^{x_0^-} dx\ e^{-i \beta \phi^-}  &        \\
     & i \int_{-L}^{x_0^-} dx\ e^{i \beta \phi^-}
\end{pmatrix}. \non\\
\ee
Notice that for $-i u \to \infty$ the leading contribution comes from the $Z^{\pm}_{11}$ elements.
This observation will be subsequently quite useful.

We shall first derive the associated local integrals of motion. In particular, the energy and momentum in the presence of defect will be obtained. Let us first recall the generating function of the local integrals of motion
\be
{\cal G}(\lambda) = \ln\ [ tr T^+(L, x_0, \lambda)\ \tilde L(x_0, \lambda)\ T^-(x_0, L, \lambda)]
\ee
Schwartz boundary conditions are imposed at the end point of the system $\pm L$. Recalling also the ansatz for the monodromy matrices
we conclude that the generating function:
\be
{\cal G}(\lambda) = \ln tr\Big [e^{Z^+(L, x_0)} (1+W^+(x_0))^{-1} (\Omega^+(x_0))^{-1} \tilde L(x_0) \Omega^-(x_0) (1+W^-(x_0))e^{Z^-(x_0, -L)}\Big ]
\ee
Choosing to consider the $-iu \to \infty$ behavior we take into account the leading contribution for the $Z^{\pm}_{11}$ terms, then the generating function takes the form reads as:
\be
{\cal G}(\lambda) = Z^{+}_{11} + Z^-_{11} + \ln \Big [(1+W^+(x_0))^{-1} (\Omega^+(x_0))^{-1} \tilde L(x_0) \Omega^-(x_0) (1+W^-(x_0))  \Big]_{11}
\ee
Expanding the latter expression in powers of $u^{-1}$ we obtain the following:
\be
{\cal G}(\lambda) = \sum_{m=0}^{\infty} {I^{(m)} \over u^m}.
\ee
Recalling now the expression for the generating function of integrals of motion we conclude that
\be
I^{(1)} &=& - {m \over 4i } \int_{-L}^{x_0^-}dx\  \Big ( -{\beta^2 \over 2m^2}{\mathfrak f}^{-2}(x) + \cos(\beta \phi^-(x)) \Big ) \non\\&-& {m \over 4i } \int_{x_0^+}^L dx\  \Big ( -{\beta^2 \over 2m^2}{\mathfrak f}^{+2}(x) + \cos(\beta \phi^+(x)) \Big ) \non\\
&+& {i\over {\cal D}} \Big ( e^{-{i\beta \over 4}(\phi^+(x_0) +\phi^-(x_0))} \bar a - e^{{i\beta \over 4}(\phi^+(x_0) +\phi^-(x_0))} a \Big )+{\beta \over 2m {\cal D}}\Big ({\mathfrak f}^+(x_0)+{\mathfrak f}^-(x_0)\Big ){\cal A} \nonumber
\ee
where we define:
\be
{\cal D} &=& e^{-{i\beta \over 4}(\phi^+(x_0) -\phi^-(x_0))}V + e^{{i\beta \over 4}(\phi^+(x_0) -\phi^-(x_0))}V ^{-1}, \non\\
{\cal A} &=& e^{-{i\beta \over 4}(\phi^+(x_0) -\phi^-(x_0))}V -e^{{i \beta \over 4}(\phi^+(x_0) -\phi^-(x_0))}V^{-1}.
\ee
If we now perform the same expansion for $\lambda \to -\infty$, we basically end up with a similar
expression, by simply exploiting the fundamental symmetry of the monodromy matrix:
\be
T(u^{-1}, \phi,\ \pi, V, a, \bar a) = T(-u,\ -\phi,\ \pi, V^{-1}, a, \bar a). \label{symm}
\ee
More precisely, one concludes that:
\be
I^{(-1)} &=& {m \over 4i } \int_{-L}^{x_0^-}dx\  \Big ( -{\beta^2 \over 2m^2}\hat {\mathfrak f}^{-2}(x) + \cos( \beta \phi^-(x)) \Big )\non\\ &-& {m \over 4i } \int_{x_0^+}^L dx\  \Big ( -{\beta^2 \over 2m^2}\hat {\mathfrak f}^{+2}(x) + \cos (\beta \phi^+(x)) \Big ) \non\\
&-& {i\over {\cal D}} \Big ( e^{{i\beta \over 4}(\phi^+(x_0) +\phi^-(x_0))} \bar a - e^{-{i\beta \over 4}(\phi^+(x_0) +\phi^-(x_0))} a \Big )+ {\beta \over 2m {\cal D}} \Big (\hat {\mathfrak f}^+(x_0) + \hat {\mathfrak f}^-(x_0) \Big ) {\cal A} \nonumber
\ee
where we define
\be
\hat {\mathfrak f}^{\pm}(\phi,\ \pi) = {\mathfrak f}^{\pm}(-\phi,\ \pi).
\ee
Any combination of the quantities $I^{(1)},\ I^{(-1)}$ can be picked as one of the
charges in involution. In particular the defect-extended form of the sine-Gordon Hamiltonian is defined as:
\be
{\cal H} &=& {2 i m \over \beta^2}(I^{(1)} -I^{(-1)}) \non\\
&=& \int_{-L}^{x_0^-} dx\ \Big ( {1\over 2} (\pi^{-2}(x) + \phi^{-'2}(x)) - {m^2 \over \beta^2}\cos(\beta \phi^-(x)) \Big ) \non\\ &+& \int_{x_0^+}^{L} dx\ \Big ( {1\over 2} (\pi^{+2}(x) + \phi^{+'2}(x)) - {m^2 \over \beta^2}\cos(\beta \phi^+(x))\Big )\non\\
&-& {4m \over \beta^2{\cal D}} \cos {\beta \over 4}( \phi^+(x_0) +
\phi^-(x_0))\ \Big (\bar a - a\Big ) +{2i \over \beta {\cal D}}
\Big (\phi^{+'}(x_0) + \phi^{-'}(x_0)\Big) {\cal A} \label{hh1} \nonumber
\ee
and we also identify the sine-Gordon momentum as:
\be
{\cal P} &=& {2im \over \beta^2} \Big (I^{(1)} +I^{(-1)} \Big )\non\\& =&
\int_{-L}^{x_0^-}dx\ \phi^{-'}(x) \pi^-(x)+ \int_{x_0^+}^{L}dx\ \phi^{+'}(x) \pi^+(x)\non\\
&+& {4mi \over \beta^2 {\cal D}} \sin{\beta \over 4}(\phi^+(x_0) + \phi^-(x_0))\ \Big (\bar a + a \Big ) +
{2i \over \beta {\cal D}} \Big ( \pi^+(x_0) + \pi^-(x_0) \Big ) {\cal A}.  \label{pp1} \nonumber
\ee

Explicit computation of the Poisson bracket $\{ {\cal H}, {\cal P} \}$ now yields a number of non-zero
terms; we shall come back to this issue after deriving the sewing conditions.

The next step is the derivation of the time components of the associated Lax pairs.
Expressions of the time component ${\mathbb V}$ of the Lax pair are known (see e.g. \cite{sts, ftbook}).
The generic expressions for the bulk left and right theories as well as the defect points are given as \cite{avan-doikou-defect, avandoikou}:
\be
&&{\mathbb V}^{+}(x, \lambda, \mu) = t^{-1}(\lambda) tr_a \Big (T_a^+(L, x ,\lambda) r_{ab}(\lambda -\mu)T_a^+(x, x_0, \lambda)
\tilde L_a(x_0, \lambda)T_a^-(x_0, -L, \lambda) \Big ) \non\\
&& {\mathbb V}^{-}(x, \lambda, \mu) = t^{-1}(\lambda) tr_a \Big (T_a^+(L, x_0 ,\lambda)\tilde L_a(x_0) T_a^-(x_0,x,\lambda)
r_{ab}(\lambda -\mu)T_a^-(x, -L, \lambda) \Big ) \non\\
&& \tilde {\mathbb V}^+(x_0, \lambda, \mu) = t^{-1}(\lambda) tr_a \Big ( T_a^+(L, x_0, \lambda) r_{ab}(\lambda-\mu)
\tilde L_a(x_0, \lambda) T_a^-(x_0,-L, \lambda)\Big )\non\\
&& \tilde {\mathbb V}^-(x_0,\lambda, \mu)= t^{-1}(\lambda) tr_a \Big (T_a^+(L, x_0, \lambda) \tilde L_a(x_0, \lambda)
 r_{ab}(\lambda -\mu)T_a^-(x_0, -L, \lambda) \Big ).
\label{timecomp}
\ee
In order to identify the Lax pair associated to the Hamiltonian and momentum
it is necessary to formulate the expansion of ${\mathbb V}$ in both negative and positive powers of $u$.

The first order contribution in the $u^{-1}$ expansion  of the bulk ${\mathbb V}^{\pm}$ operator (we have set the second spectral parameter $v \equiv e^{\mu} $) reads:
\be
{\mathbb V}^{\pm(1)} = {\beta^2 \over 8}\left ({\beta \over 2m}\sigma^z (\pi^{\pm} + \phi^{\pm'}) +i v \Big (\sigma^-e^{-{i \beta \over 2} \phi^{\pm}} -
\sigma^+ e^{{i \beta \over 2} \phi^{\pm}} \Big ) \right ) \label{V1}
\ee
The first order contribution in the $u$ expansion reads:
\be
\hat {\mathbb V}^{\pm(1)} =  {\beta^2 \over 8}\left ({\beta \over 2m}\sigma^z (\pi^{\pm} - \phi^{\pm '}) -i v^{-1} \Big (\sigma^-e^{{i \beta \over 2} \phi^{\pm}} -
\sigma^+ e^{-{i \beta \over 2} \phi^{\pm}} \Big ) \right ) \label{v2}
\ee

Subtracting these two expressions and multiplying by $-{2i m\over \beta^2}$ we obtain the time component
of the Lax pair associated to the Hamiltonian:
($\Omega^{\pm} = e^{{i\beta \over 4} \phi^{\pm} \sigma^z}$)
\be
{\mathbb V}^{\pm}_{{\cal H}} = {\beta \over 4i } \phi^{\pm '} \sigma^z +{v m \over 4 i } \Omega^{\pm}\sigma^y(\Omega^{\pm})^{-1} + {v^{-1}m \over 4 i } (\Omega^{\pm})^{-1}\sigma^y\Omega^{\pm}
\ee
Adding now (\ref{V1}), (\ref{v2}), after multiplying with $-{2i m\over \beta^2}$, provides the time component
of the Lax pair associated to the momentum:
\be
{\mathbb V}^{\pm}_{{\cal P}} = {\beta \over 4i } \pi^{\pm} \sigma^z +{vm \over 4 i } \Omega^{\pm}\sigma^y(\Omega^{\pm})^{-1} - {v^{-1} m\over 4 i } (\Omega^{\pm})^{-1}\sigma^y \Omega^{\pm}
\ee

The next step is the derivation of the relevant Lax pairs for the defect point from the left and
the right, based on the expression (\ref{timecomp}). Indeed, after some cumbersome but
quite straightforward computations, and after defining:
\be
w^{\pm} = -{i \beta \over m} {\mathfrak f}^{\pm}, ~~~~~\hat w^{\pm} = {i\beta\over m}\hat {\mathfrak f}^{\pm},
\ee
we conclude from the expansion in powers of $u^{-1}$:
\be
{\tilde {\mathbb V}}^{+(1)} &=&  {i\beta^2 \over 8} {\cal D}^{-2}\sigma^z \Big [w^+  + w^- + e^{{i \beta \over 2} \phi^-} V a+e^{-{i \beta \over 2} \phi^-} V^{-1} \bar a \Big ]  \non\\
&+& {i \beta^2 \over 4}{\cal D}^{-1} v \Big [\sigma^-e^{-{i \beta \over 4}(\phi^+ + \phi^-)}V^{-1}  -\sigma^+e^{{i\beta \over 4}(\phi^+ +\phi^-)}V \Big ],
\ee
whereas the expansion in powers of $u$ leads to:
\be
\hat{\tilde {\mathbb V}}^{+(1)} &=& -{i\beta^2 \over 8} {\cal D}^{-2} \sigma^z \Big [\hat w^+   + \hat w^- - e^{{i \beta \over 2} \phi^-}V \bar a- e^{-{i \beta \over 2}\phi^-} V^{-1} a  \Big ] \non\\
&-& {i \beta^2 \over 4}{\cal D}^{-1} v^{-1} \Big [\sigma^-e^{{i \beta \over 4}(\phi^+ + \phi^-)}V  -\sigma^+e^{-{i\beta \over 4}(\phi^+ + \phi^-)}V^{-1} \Big ]
\ee
Similarly, the corresponding expressions for $\tilde {\mathbb V}^{-(1)},\ \tilde {\mathbb V}^{-(1)}$ are given below:
\be
{\tilde {\mathbb V}}^{-(1)} &=&  {i\beta^2 \over 8}{\cal D}^{-2}\sigma^z \Big [w^+  + w^- - e^{{i \beta \over 2}\phi^+ }V^{-1} a -e^{-{i \beta \over 2}\phi^+ } V \bar a \Big ] \non\\
&+&  {i \beta^2 \over 4} {\cal D}^{-1}v \Big [\sigma^-e^{-{i \beta \over 4}(\phi^+ + \phi^-)}V  -\sigma^+e^{{i\beta \over 4}(\phi^+ + \phi^-)}V^{-1} \Big ]
\ee
\be
\hat{\tilde {\mathbb V}}^{-(1)} &=-& {i\beta^2 \over 8} {\cal D}^{-2} \sigma^z \Big [\hat w^+ + \hat w^- +e^{{i \beta \over 2}\phi^+}V^{-1}\bar a +e^{-{i \beta \over 2}\phi^+ }V a \Big ] \non\\ &-&  {i \beta^2 \over 4} {\cal D}^{-1} v^{-1} \Big [\sigma^-e^{{i \beta \over 4}(\phi^+ + \phi^-)}V^{-1} -\sigma^+e^{-{i\beta \over 4}(\phi^+ + \phi^-)}V \Big ].
\ee

We are now in a position to apply the scheme elaborated in \cite{avan-doikou-defect}.
The first manifest observation from the continuity conditions
\be
&& \tilde {\mathbb V}^{+(1)}(x_0) \to {\mathbb V}^{+(1)}(x_0^+), ~~~~~x_0^+ \to x_0 \non\\
&& \tilde {\mathbb V}^{-(1)}(x_0) \to {\mathbb V}^{-(1)}(x_0^-), ~~~~~x_0^- \to x_0  \label{cont}
\ee
(similar continuity conditions apply for the ``hatted'' quantities, but are omitted for brevity),
is that:
\be
V= e^{{i\beta \over 4}(\phi^+ - \phi^-)}. \label{sew1a}
\ee
and will be hereafter denoted as ``first sewing condition $S_1$ ''. Remember that from the very beginning
one has already fixed the Casimir $C_0$ to some value $c_0$ independently of any sewing requirement. This can be seen
as an ``order zero condition $S_0$ '' without any dependance in the bulk variables and yields a first-class
constraint Poisson-commuting with all dynamical variables.

After imposing (\ref{sew1a}) the time components of the Lax
pairs on the defect point take the following simple expressions:
\be
\tilde {\mathbb V}^{\pm(1)} = {\beta^2 \over 8} \left ( {\beta \over 4m}\sigma^z \Big (\pi^+ + \phi^{+'} + \pi^-+\phi^{-'}) \pm {i\sigma^z\over 4}  {\mathbb M} + i v \Big (\sigma^-e^{-{i \beta \over 2} \phi^{\pm}} - \sigma^+ e^{{i \beta \over 2} \phi^{\pm}} \Big ) \right )
\ee
and the first term in the $u$ expansion provides:
\be
\hat{\tilde {\mathbb V}}^{\pm} = {\beta^2 \over 8} \left ({\beta \over 4 m}\sigma^z \Big (\pi^+ - \phi^{+'} + \pi^- -\phi^{-'}) \pm {i\sigma^z\over 4}  \hat {\mathbb M} - i v^{-1} \Big (\sigma^-e^{{i \beta \over 2} \phi^{\pm}} - \sigma^+ e^{-{i \beta \over 2} \phi^{\pm}}\Big )\right )
\ee
where we define:
\be
&& {\mathbb M} = e^{-{i\beta \over 4}(\phi^+ + \phi^-)}\bar a + e^{{i\beta \over 4}(\phi^+ + \phi^-)}a \non\\
&& \hat {\mathbb M} = e^{{i\beta \over 4}(\phi^+ + \phi^-)}\bar a + e^{-{i\beta \over 4}(\phi^+ + \phi^-)}a
\ee
Continuity conditions on the Lax pair as also described in (\ref{cont})
give rise to the following sewing conditions on the defect point $x_0$ associated to the momentum and the Hamiltonian respectively:
\be
S_2:~~~~&& \pi^{+}(x_0) - \pi^{-}(x_0) = {i m \over \beta} \cos {\beta \over 4} (\phi^+(x_0) +
\phi^-(x_0))\ \Big (a +\bar a \Big )\non\\
S'_2:~~~~&& \phi^{+'}(x_0)- \phi^{-'}(x_0) =  {m \over \beta} \sin {\beta \over 4} (\phi^+(x_0) +
\phi^-(x_0))\ \Big (\bar a - a \Big ) \label{sew2}
\ee
the prime denotes the derivative with respect to $x$.

It is instructive to point out that
comparison of the extracted charges (\ref{hh1}), (\ref{pp1}), and the latter equations
(\ref{sew2}) with similar results obtained for instance in \cite{haku, caudr} reveal
manifest discrepancies. We shall further comment on this matter in the discussion section.

Consistency of the sewing conditions $S_1,\ S_2,\ S'_2$ can now be checked by computing their Poisson
brackets with the first two Hamiltonians ${\cal H}, {\cal P}$. Indeed one gets:
\be
\Big \{ {\cal H}, S_1 \Big \} = -\frac{i\beta}{4} (\pi^{+}(x_0) - \pi^{-}(x_0)) S_1 &+&
\frac{i\beta}{4} S_2 V + o({\cal D} -2)\non\\
\Big \{ {\cal P}, S_1 \Big \} = -\frac{i\beta}{4} (\phi^{+'}(x_0)- \phi^{-'}(x_0)) S_1 &+&
\frac{i\beta}{4} S'_2 V + o({\cal D} -2)\label{PBc1}
\ee
We recall that on-shell ${\cal D} \approx 2$ ; ${\cal A} \approx 0$.

Consider now the Poisson brackets of ${\cal H},\ {\cal P}$ with $S_2,\ S'_2$. One easily obtains that
they are given by expressions of the following form:
\be
&& \Big \{{\cal P}, S_2 \Big \} = (\pi^{+'}(x_0) - \pi^{-'}(x_0))\
F(\pi^{+}(x_0) + \pi^{-}(x_0), \phi^{+'}(x_0)+ \phi^{-'}(x_0),  \phi^{+}(x_0), \phi^{-}(x_0), V,a,\bar a)
\non\\
&& \Big \{{\cal H}, S_2 \Big \} =(\phi^{+''}(x_0)- \phi^{-''}(x_0))\
G(\pi^{+}(x_0) + \pi^{-}(x_0), \phi^{+'}(x_0)+ \phi^{-'}(x_0),  \phi^{+}(x_0), \phi^{-}(x_0), V,a,\bar a) \non\\
\label{PBc2}
\ee
where $F$ and $G$ are given functions to be computed specifically. Poisson brackets with $S'_2$ are given
by similar expressions exchanging ${\cal H}$ and $ {\cal P}$. Note that (contrary to the
non-linear Schroedinger case) no term proportional to the singular contribution $\delta (0)$ arise,
they fully cancel in the Poisson brackets. It is therefore to be expected that the finite terms
on the r.h.s. of both PB's will yield the third sewing conditions $S_3, S'_3$ which will respectively take the
form (expected from general arguments)
$(\pi^{+'}(x_0) - \pi^{-'}(x_0)) = -F$ and $(\phi^{+''}(x_0)- \phi^{-''}(x_0) = -G$. Explicit derivation
of these sewing conditions from higher terms in the expansion of the $\\{\mathbb V}$ operators is technically
quite cumbersome but we conjecture that they will coincide with the rhs of (\ref{PBc2}).

Let us now reconsider the Poisson bracket $\{ {\cal H}, {\cal P} \}$. It turns out from explicit
computations that one has in fact:

\be
\Big \{ {\cal H},\ {\cal P} \Big \} \approx 0,
\ee
i.e. the Poisson bracket vanishes provided that the constraints $S_1,\ S_2,\ S'_2$ be satisfied.

Assuming
that the Hamiltonians ${\cal H}$ and $ {\cal P}$ weakly preserve all constraints (as already
established for $S_1$ and conjectured for $S_2, S'_2$ ) we deduce that the momentum and Hamiltonian Dirac commute.
Our construction of Type II defect is thus compatible with a statement of Liouville-integrability on-shell. A general formal argument based on the underlying Poisson structure has developed in \cite{avan-doikou-defect} establishing the property of weak preservation of constraints.
\\
\\
{\bf Equations of motion}\\
To extract the associated equations of motion for the left and right bulk theories as well as the defect point
one needs to employ the zero curvature condition expressed as:
\be
\dot{\mathbb U}^{\pm}(x,t) - {\mathbb V}^{\pm'}(x,t) + \Big [{\mathbb U}^{\pm}(x, t), {\mathbb V}^{\pm}(x, t) \Big ] =0 ~~~~~x \neq x_0.
\label{zero1}
\ee
As usual the dot denotes derivative with respect to $t$.

On the defect point in particular the zero curvature condition is formulated as (this is also
transparent when discussing the continuum limit of discrete theories (see e.g. \cite{avan-doikou-defect})
\be
\dot {\tilde L}(x_0)  = \tilde {\mathbb V}^+(x_0) \tilde L(x_0) -
\tilde L(x_0) \tilde {\mathbb V}^{-}(x_0), \label{zerod}
\ee
and describes explicitly the jump occurring across the defect point.

The equations of motion are obtained via the zero curvature conditions as described above
or (equivalently thanks to the sewing conditions) via the Hamiltonian equations i.e.
\be
&& \dot \phi^{\pm} = \Big \{ {\cal H},\ \phi^{\pm}\Big \}, ~~~~~\dot \pi^{\pm} = \Big \{ {\cal H},\ \pi^{\pm}\Big \}, \non\\
&& \dot {\mathrm e} = \Big \{ {\cal H},\ {\mathrm e} \Big \}, ~~~~~{\mathrm e} \in \Big\{a,\ \bar a,\  V \Big \}
\ee
bear also in mind that
\be
\Big \{ \pi^{\pm},\ {\mathrm e} \Big \} = \Big \{ \phi^{\pm},\ {\mathrm e}\Big \} =0.
\ee

For the left and right bulk theories the familiar equations of motion for the sine-Gordon model arise
\be
\ddot{\phi}^{\pm}(x, t) - \phi^{\pm''}(x, t) + {m^2 \over \beta} \sin ( \beta \phi^{\pm}(x, t))=0
\ee
On the defect point the time evolution of the defect degrees of freedom are obtained as:
\be
\dot a &=&  -{m \over 2 {\cal D}^{2}}\  {\cal A}\ a\ \cos {\beta \over 4} (\phi^+ + \phi ^-)\ \Big (\bar a - a \Big )
- {m \over {\cal D}}\ \cos{\beta \over 4}(\phi^+ + \phi^-)\ \Big (V^2 -V^{-2}\Big ) \non\\
&-& {\beta i \over  {\cal D}^2}\ a\ \Big (\phi^{+'} + \phi^{-'}\Big  )
\ee
\be
\dot{\bar a} &=& {m \over 2 {\cal D}^2}\ {\cal A}\ \bar a\ \cos{\beta \over 4}(\phi^+ + \phi^-)\ \Big (\bar a - a\Big ) - {m \over  {\cal D}}\ \cos{\beta \over 4}(\phi^+ + \phi^-)\ \Big (V^2 -V^{-2}\Big ) \non\\
&+& {i \beta \over  {\cal D}^2}\ \bar a\  \Big ( \phi^{+'} + \phi^{-'} \Big )
\ee
\be
\dot V = {m \over 2 {\cal D}}\ V\ \cos ({\beta \over 4} (\phi^+ + \phi^-))\  \Big (a +\bar a \Big ).
\ee

With this we conclude our presentation on the sine-Gordon model in the presence of integrable dynamical defects.

\subsection*{Acknowledgments}
Material of this article was also presented by A. Doikou in
``Integrable systems and Quantum Symmetries'', Prague, June 2012.

\end{document}